# Fast switchable electro-optic radial polarization retarder

**B. C. Lim,[1,2] P. B. Phua,[3,4*] W. J. Lai[4] and M. H. Hong[1,2]**

[1]*Data Storage Institute, (A*STAR), DSI Building, 5, Engineering Drive 1, Singapore 117608*

[2]*Department of Electrical and Computer Engineering, National University of Singapore, 4, Engineering Drive 3, Singapore 117576*

[3]*DSO National Laboratories, 20, Science Park Drive, Singapore 118230*

[4]*Nanyang Technological University, 50, Nanyang Avenue, Singapore 639798*

[*]*Corresponding author: ppohboon@alum.mit.edu*

A fast, switchable electro-optic radial polarization retarder (EO-RPR) fabricated using the electro-optic ceramic PMN-PT is presented. This EO-RPR is useful for fast, switchable generation of pure cylindrical vector beam. When used together with a pair of half-wave plates, the EO-RPR can change circularly polarized light into any cylindrical vector beam of interest such as radially or azimuthally polarized light. Radially and azimuthally polarized light with purities greater than 95% are generated experimentally. The advantages of using EO-RPR include fast response times, low driving voltage and transparency in a wide spectral range (500 -7000 nm). © 2008 Optical Society of America

Transparent electro-optic (EO) ceramics have widespread applications such as variable optical attenuator, polarization controllers, Q-switches, flash goggles, colour filters and high power ceramic lasers [1-3]. Recently, Jiang et al. [1] presented a newly developed transparent EO ceramic, Pb(Mg$_{1/3}$Nb$_{2/3}$)O$_3$-PbTiO$_3$ (PMN-PT) which is transparent from 500 to 7000 nm. It belongs to a class of ferroelectric material called relaxor ferroelectrics. This class of materials exhibits minimal hysteresis (desirable in any device application), resolving the issue of high hysteresis experienced by the EO ceramic PLZT. PMN-PT also has a lower field induced optical loss (which degrades device performance) than PLZT and boasts an EO effect about 2 to 5 times larger than that of PLZT and nearly 100 times better than that of LiNbO$_3$. This significantly lowers the applied voltage requirement, to achieve the same phase shift, for PMN-PT compared to PLZT or LiNbO$_3$. Compared to using single crystal materials like LiNbO$_3$, no consideration of the crystalline orientation is required when using PMN-PT, as it is polycrystalline and thus polarization independent. The mature hot-pressing technique also makes PMN-PT cheaper and easier to fabricate than single crystalline materials. Furthermore, PMN-PT has a fast EO response time in the sub-microsecond region compared to the millisecond response of liquid crystals. This is due to the much faster electronic and ionic movements in PMN-PT compared to molecular movement in liquid crystals. It had been demonstrated in [1] that the switching speed of such EO thin film ceramic is 10 ns while that of the bulk EO ceramic is less than 100 ns.

In this paper, a polarization optical element called electro-optic radial polarization retarder (EO-RPR) is presented. It is fabricated using PMN-PT (thickness of 0.5 mm) from Boston Applied Technologies, to harness the material's beneficial properties as mentioned above. The EO-RPR is useful in generating cylindrical vector beams in a switchable manner. Such beams have cylindrical symmetry in polarizations and are useful, especially the radially polarized light, in numerous applications. Examples include a tighter focal spot which is desirable in optical data storage [4], optical trapping of metallic particles [5] and measurement probe optimization for localized surface plasmon microscopy [6]. Cylindrical vector beams can also be used to enhance laser cutting of metals [7] and for focus shaping [8]. Numerous schemes have been proposed for generating cylindrical vector beams. Traditionally, Tidwell et al. combined two orthogonally polarized beams interferometrically [9] while other more recent methods have employed specially designed polarization elements for generation [10-13]. Compared to previous methods, our EO-RPR adds a fast switching ability to the generation of pure cylindrical vector beams, all done without the need to move or change any optical components. It is also transparent over a wide spectral range from visible to infrared (500-7000 nm) and easily fabricated as PMN-PT is easily processed by laser cutting and drilling. An example of an application that requires the switchability of our EO-RPR is the recently reported second-harmonic-generation microscopy for the determination of three-dimensional orientation of molecules [14]. In [14], the authors used an eight-segment polarization converter, based on liquid crystals, which has a response time of ~150 ms and generates only pseudo-radial polarization due to the segmented polarization conversions. These can be improved by using our EO-RPR which gives faster switching and more accurate radial polarization profile. The response time of PMN-PT is in the sub-microsecond region as mentioned previously [1] and the setup of the EO-RPR will produce pure radial polarization as presented hereinafter.

Fig. 1a shows the schematic of the EO-RPR. A Q-switch, diode-pumped solid state (DPSS) 355 nm ultra-violet laser is used to cut the PMN-PT substrate into a disc (diameter of 5 mm) followed by laser drilling of a small through hole at the center of the disc. A copper wire (diameter of 60 μm) acting as the positive electrode is then inserted into the hole. The disc is in turn placed inside a central through hole (diameter of 5 mm) of an aluminium holder which acts as the negative electrode. According to [15], PMN-PT is optically isotropic in the absence of an electric field but becomes anisotropic in the presence of an electric field. The refractive index is lower in the direction of the applied field compared to that perpendicular to the field. Due to this arrangement of electrodes, when a voltage is applied between the copper wire and the aluminium holder, a radial electric field is formed across the PMN-PT disc, resulting in a polarization retarder with radial fast axes, namely a radial polarization retarder. PMN-PT relies on the quadratic Kerr electro-optics effect instead of Pockels electro-optics effect [1]. Its electrically induced phase retardation is given by equation (1), where $V$ is the applied voltage between electrodes, d is the distance between electrodes, $\lambda$ is the wavelength, $\Delta\theta$ is the polarization retardation angle, $L$ is the thickness of the ceramic, $r$ is the Kerr coefficient of the ceramic and $n_0$ is the ceramic's refractive index in the absence of an electric field.

$$\Delta\theta = \frac{\pi n_0^3 L r V^2}{\lambda d^2}. \tag{1}$$

Fig. 1b shows the schematic of the experimental setup. For convenience, a frequency doubled DPSS CW laser operating at 532 nm is used as the laser source. A pure linearly polarized $LG_{01}$ mode of $M^2 \sim 2$ is first generated using a Spiral Phase-Plate (SPP) followed by spatial filtering as shown in the red dotted box. A circular polarizer (which comprises of a linear polarizer followed by a quarter wave-plate) is then used to convert its linear polarization into circular polarization. This circularly polarized $LG_{01}$ mode serves as the input light source for the EO-RPR. To ensure the quality of the generated cylindrical vector beam, alignment of the center of the EO-RPR to that of the incident light is essential. Voltage is applied to the EO-RPR to produce a quarter-wave retardation. The calculated required voltage is 495 V using the following values $\Delta\theta = \pi/2$, $\lambda = 532$ nm, $L = 0.5$ mm, $r = 8.69 \times 10^{-16}$ m$^2$/V$^2$, $n_0 = 2.5$ and $d = 2.5$ mm in equation (1). Fig. 2a shows the polarization conversions for incident right-hand and left-hand circular polarizations. The polarization conversions can be visualized using a Poincaré sphere [16] as shown in Fig. 2b.

Any polarization state change is represented on the Poincaré sphere as a clockwise rotation about the fast axis of the polarization component, with the rotation angle given by its retardation angle. To illustrate, for right-hand circularly polarized light incident on a standard quarter-wave plate (retardation angle of $\pi/2$) with a horizontal fast axis, the polarization state change can be traced by the red dotted arrow (1) as a clockwise rotation about the positive axis of $s_1$ from right-hand circular polarization to 45° linear polarization. If the quarter-wave plate is now rotated such that its fast axis is vertical, the polarization state change for an incident right-hand circularly polarized light is then traced by the red dotted arrow (3) ($\pi/2$ clockwise rotation about the negative axis of $s_1$) and the output polarization is 135° linearly polarized. On this basis, dotted arrows (2) and (4) correspond to polarization state changes for an incident right-hand circularly polarized light through a standard quarter-wave plate with fast axes at 45° and 135° from the horizontal respectively. Now consider a right-hand (or left-hand) circularly polarized light incident on the EO-RPR (with their centers overlapped). Since the EO-RPR has a continuously varying radial fast axes, it can be visualized that at any angular position $\theta$ (also the orientation of the fast axis) along the radius of the cross section of the light beam, the output polarization state will be a point on the equator of the Poincaré sphere, and, in physical space, the output polarization will be a linear polarization with orientation of $\theta + 45°$ (or $\theta - 45°$ in the case of incident left-hand circular polarization). Thus, generating a pure cylindrical vector beams as shown in Fig. 2a.

The exiting light from the EO-RPR is subsequently passed through a pair of half-wave plates which work in tandem to rotate polarization states R or L to any cylindrical vector beams of interest including radially or azimuthally polarized light [8]. The far field intensity of the generated cylindrical vector beam is then imaged using a CCD camera placed at the focal plane of a lens. Note that due to the nature of polarization conversion adopted, a uniform beam exiting the EO-RPR will accumulate a spiral phase front. This spiral phase front is due to Pancharatnam's phase as reported in [13]. In our case, we have compensated this spiral phase front by the initial conversion of the laser light into a $LG_{01}$ mode of the opposite spiral phase using a SPP.

Radially polarized light is generated experimentally using the aforementioned proposed scheme. Fig. 3 shows the observed far field intensity distribution after fine tuning the applied voltage to 520 V. The higher voltage used compared to that calculated might be due to the slightly larger than expected size of the central hole of the aluminium holder. From Fig. 3, it can be seen that a clear doughnut shape is obtained in the far field. Note that this good radially polarized donut beam is generated without any spatial filtering. A polarizer is used to check the polarization state of the light generated and two lobes are seen rotating with the rotated polarizer. The white arrows shown indicate the direction of the transmission axis of the rotated polarizer and it can be deduced that the generated light is indeed radially polarized. Using a 2D polarimeter, polarization purity of the generated light is measured. Polarization purities of 95% and 97% are obtained for the generated radially and azimuthally polarized light respectively. The $M^2$ value of the generated light is 2.55 as compared to the initial value of 2 for the incident $LG_{01}$ mode.

In conclusion, transparent EO ceramic PMN-PT was used for the fabrication of EO-RPR which is useful for generating switchable cylindrical vector beams. It was demonstrated experimentally that this scheme can generate radially and azimuthally polarized light with polarization purities larger than 95%. The unique feature of this scheme as compared to previous methods of cylindrical vector beam generation is its potential for fast, switchable generation. Potential applications which may find this fast switchability useful include Q-switching for radially polarized lasers, controlled particle acceleration and second-harmonic-generation microscopy. Furthermore, the EO-RPR can be used at various wavelengths by adjusting the applied voltage according to the wavelength of interest. Also, the EO-RPR can be easily fabricated as PMN-PT is easily cut using a Q-switch DPSS 355 nm laser and no nano-size features are required.

## Acknowledgements

This work was supported by DSO National Laboratories and Data Storage Institue, A*STAR (Agency for Science, Technology and Research), Singapore. The authors acknowledge Tiaw Kay Siang, Tan Li Sirh and Dr. Larry Yuan Xiaocong for the provision of necessary optics and equipments used during experiments.

**References**

1. H. Jiang, Y. K. Zou, Q. Chen, K. K. Li, R. Zhang and Y. Wang, "Transparent electro-optic ceramics and devices," Proc. SPIE **5644**, 380-394 (2005)

2. A. J. Moulson and J. M. Herbert, "Electro-optic ceramics," in *Electroceramics: Materials, Properties, Applications.* 2$^{nd}$ Edition. (John Wiley & Sons, 2003), pp. 433-468.

3. X. S. Chen, K. W. Li, K. Zou, R. Zhang, H. Jiang, G. Ozen and B. Di Bartolo, "Novel electro-optic ceramic materials for microchip and high power lasers," in *MRS Proceedings* **782**, (2004), pp. A5.57.1-A5.57.6.

4. S. Quabis, R. Dorn, M. Eberler, O. Glöckl and G. Leuchs, "Focusing light to a tighter spot," Opt. Commun. **179**, pp. 1-7 (2000).

5. Q. Zhan, "Trapping metallic Rayleigh particles with radial polarization," Opt. Exp. **12**, pp. 3377-3382 (2004).

6. K. Watanabe, N. Horiguchi and H. Kano, "Optimized measurement probe of the localized surface plasmon microscope by using radially polarized illumination", Appl. Opt. **46**, pp. 4985-4990 (2007).

7. M. Meier, V. Romano and T. Feurer, "Materials processing with pulsed radially and azimuthally polarized laser radiation," Appl. Phys. A **86**, pp. 329-334 (2007).

8. Q. Zhan and J. R. Leger, "Focus shaping using cylindrical vector beams," Opt. Exp. **10**, pp. 324-331 (2002).

9. S. C. Tidwell, D. H. Ford and W. D. Kimura, "Generating radially polarized beams interferometrically," Appl. Opt. **29**, pp. 2234-2239 (1990).

10. H. Ren, Y. H. Lin and S. T. Wu, "Linear to axial or radial polarization conversion using a liquid crystal gel," Appl. Phys. Lett. **89**, pp. 051114-1 - 051114-3 (2006).

11. G. Machavariani, Y. Lumer, I. Moshe, A. Meir and S. Jackel, "Efficient extracavity generation of radially and azimuthally polarized beams," Opt. Lett. **32**, pp. 1468-1470 (2007).

12. P. B. Phua, W. J. Lai, Y. L. Lim, K. S. Tiaw, B. C. Lim, H. H. Teo and M. H. Hong, "Mimicking Optical Activity for Generating Radially Polarized Light," Opt. Lett. **32**, pp. 376-378 (2007).

13. Z. Bomzon, G. Biener, V. Kleiner and E. Hasman, "Radially and azimuthally polarized beams generated by space-variant dielectric subwavelength gratings," Opt. Lett. **27**, pp. 285-287 (2002)

14. K. Yoshiki, K. Ryosuke, M. Hashimoto, N. Hashimoto and T. Araki, "Second-harmonic-generation microscope using eight-segment polarization-mode converter to observe three-dimensional molecular orientation," Opt. Lett. **32**, pp. 1680-1682 (2007)

15. K. K. Li, Y. Lu and Q. Wang, US Patent 6,890,874 B1 (2005)

16. D. Goldstein, *Polarized Light*, Second Edition, (Marcel and Dekker, Inc., 2003)

**List of figure captions**

Fig. 1. (Color online) (a) Schematic of electro-optic radial polarization retarder. (b) Schematic of experimental setup.

Fig. 2. (Color online) (a) Cylindrical vector beams generated using electro-optic radial polarization retarder when incident light is circularly polarized. (b) Polarization state changes (dotted arrows) through the electro-optic radial polarizer for an incident light with right-hand circular polarization at various angular positions (1) $\theta = 0°$, (2) $\theta = 45°$, (3) $\theta = 90°$ and (4) $\theta = 135°$ of the light beam cross section.

Fig. 3. Far-field intensity distribution of radially polarized light generated using the proposed scheme. White arrows indicate the direction of the transmission axis of the rotated polarizer.

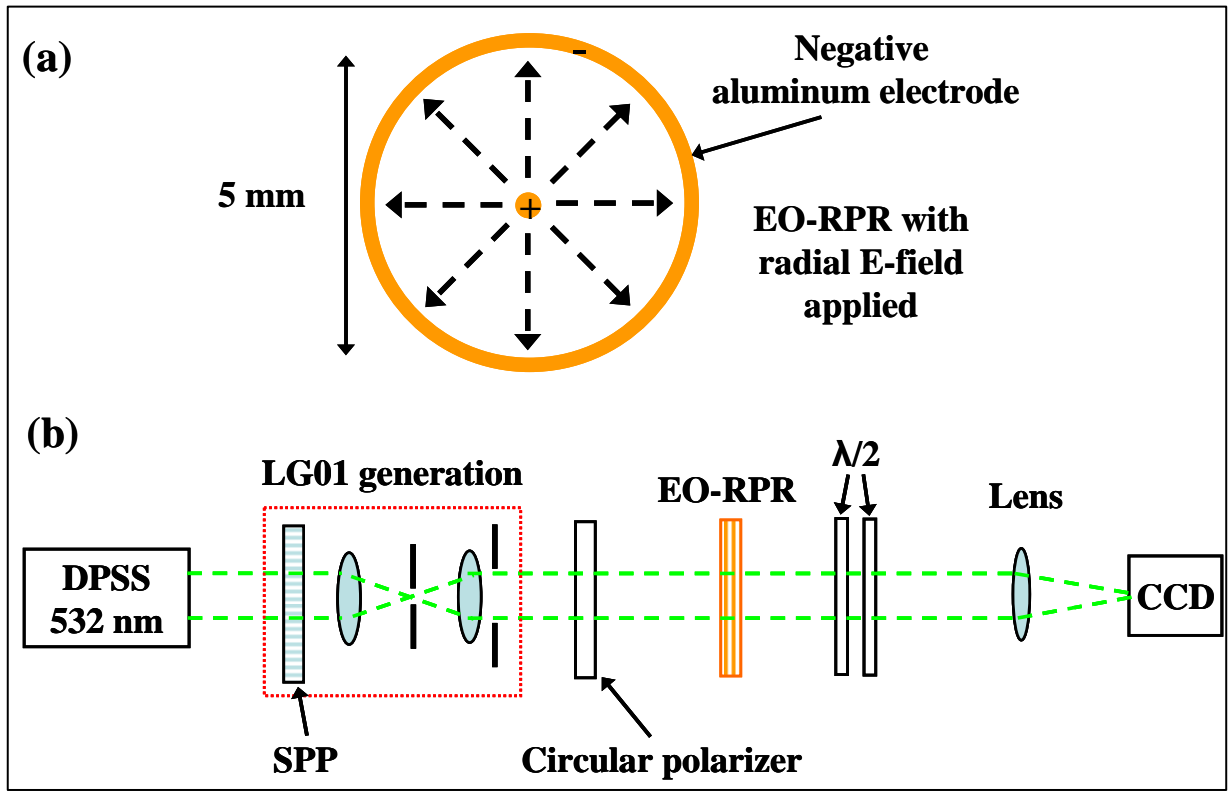

**Figure 1**

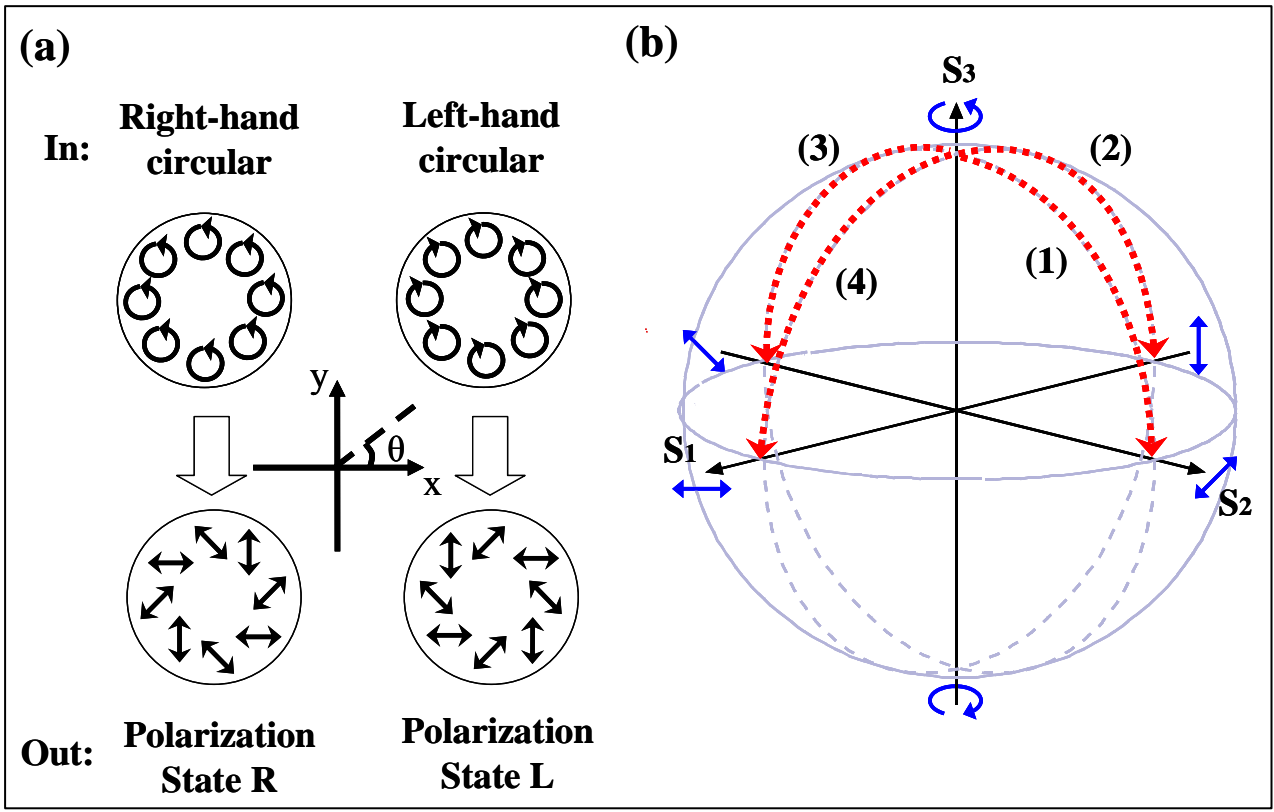

**Figure 2**

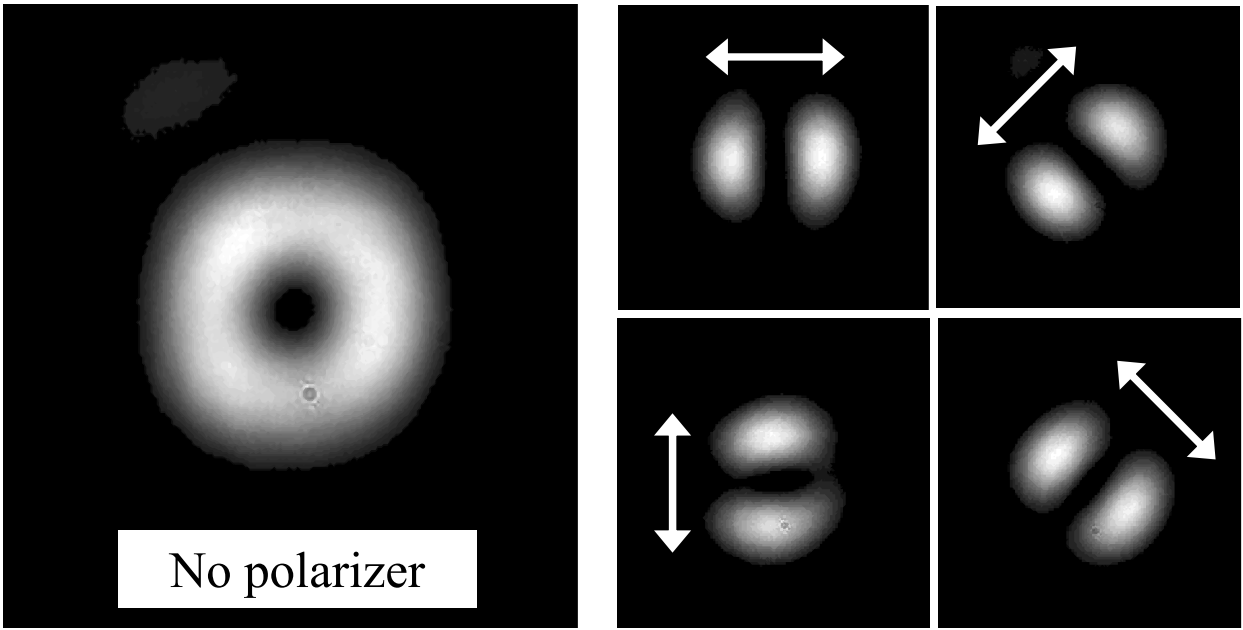

**Figure 3**